\documentclass[fleqn,usenatbib,useAMS]{mnras}
\newcommand\lsim{\mathrel{\rlap{\lower4pt\hbox{\hskip1pt$\sim$}}
        \raise1pt\hbox{$<$}}}
\newcommand\gsim{\mathrel{\rlap{\lower4pt\hbox{\hskip1pt$\sim$}}
        \raise1pt\hbox{$>$}}}
\newcommand{\msun}{{\rm M_{\odot}}}
\newcommand{\enzo}{{\sc enzo}}
 
\newcommand{\hyd}{${\rm H_2}$} 
\newcommand{\jcrit}{$J_{\rm crit}$}

\usepackage{amsmath}
\usepackage{array}
\usepackage{appendix}

\usepackage{graphicx}
\usepackage{fixltx2e}
\usepackage{booktabs}
\usepackage{tikz,xcolor,hyperref}

\definecolor{lime}{HTML}{A6CE39}
\DeclareRobustCommand{\orcidicon}{%
  \begin{tikzpicture}
    \draw[lime, fill=lime] (0,0) 
    circle [radius=0.16] 
    node[white] {{\fontfamily{qag}\selectfont \tiny ID}};
    \draw[white, fill=white] (-0.0625,0.095) 
    circle [radius=0.007];
    \end{tikzpicture}
  \hspace{-2mm}
}

\foreach \x in {A, ..., Z}{%
  \expandafter\xdef\csname orcid\x\endcsname{\noexpand\href{https://orcid.org/\csname orcidauthor\x\endcsname}{\noexpand\orcidicon}}
}

\def\apjl{ApJL}               
\def\apjs{ApJS}               
\def\aap{A\&A}                

\title[HI--Shielding of $H_2$ in Protogalaxies]{HI--shielding of ${\bf H_2}$ in UV--irradiated protogalaxies: suppression of the photodissociation rate}

\author[M. Neyer et al.]
{Meredith Neyer$^{1,2}$\thanks{E-mail: mneyer@mit.edu;
jemma@ucsb.edu}\orcidA and Jemma Wolcott-Green$^{1}$\orcidB\\
$^{1}$Department of Physics, University of California Santa Barbara, 
MC 9530, Santa Barbara, CA 93106, USA\\
$^{2}$Department of Physics and Kavli Institute for Astrophysics and Space Research, Massachusetts Institute of Technology,\\
Cambridge, MA 02139, USA}

\begin{document}

\date{}

\pagerange{\pageref{firstpage}--\pageref{lastpage}} \pubyear{2022}

\maketitle

\label{firstpage}

\begin{abstract}
We study the impact of neutral hydrogen absorption on \hyd~photodissociation
in protogalactic haloes exposed to soft-UV radiation. Lyman-series absorption
can significantly deplete dissociating photons as line overlap with the 
\hyd~Lyman-Werner bands occurs for neutral column densities exceeding 
$10^{22}~{\rm cm^{-2}}$, but this effect has not been previously included in
studies of protogalactic haloes. We use high--resolution three--dimensional
hydrodynamic simulations to investigate this ``HI--shielding'' in three
metal--free atomic cooling haloes collapsing at redshift $z\sim 10-20$. 
We use {\sc cloudy} modeling to update a previous fitting formula for 
HI–shielding which is a better model for shielding of non–ground state 
\hyd~rovibrational populations and implement the new fit in our simulations. 
We find that the inclusion of HI--shielding increases the ``critical flux'' for suppression of \hyd~cooling in these haloes 
by $\sim 60-100$ per cent. The larger critical flux has implications in particular 
for the predicted numbers of candidate haloes in which``direct collapse'' could 
seed massive ($\sim 10^5~{\rm M_\odot}$) black holes at $z\sim15$. 

\end{abstract}


\section{Introduction}
Molecular hydrogen, \hyd, has been extensively studied in the context of
the first generation of stars and galaxies, in which it plays a crucial 
role as the primary coolant of primordial gas below $\sim 10^4$K 
\citep[for a review, see][]{BYReview11}. Prior to the production 
and dispersion of metals by the supernovae, the thermodynamic evolution 
of pristine primordial gas depends sensitively on the \hyd~abundance and 
therefore on the photodissociation of \hyd, which occurs in the presence
of soft UV photons in the ``Lyman--Werner'' (LW) bands (11.1-13.6 eV).  

Depletion of \hyd~by LW radiation has been shown to raise the minimum 
mass of protogalactic haloes in which gas is able to condense and cool, 
thus delaying star formation in smaller ``minihaloes'' 
\citep{HRL97,HAR00,MBA01,Yosh+03,MBH06,WA07,ON08,Kulkarni+20, Schauer+21}. 
In more massive haloes, with virial temperatures $\gsim 10^4$K, cooling 
by neutral hydrogen allows gas to condense in haloes even in the absence 
of significant \hyd~cooling, rendering these ``atomic cooling haloes'' 
(ACHs) less vulnerable to feedback from a cosmological background LW 
radiation \citep[e.g.][]{OH02}. Typically, the column densities of 
\hyd~in ACHs grow large enough that \hyd~becomes self--shielding: 
systematic depletion of LW band photons in the outer layers of the halo 
depresses photodissociation of \hyd~in the core, allowing the gas to cool 
to temperatures of a few hundred Kelvin. However, sufficiently strong 
LW radiation fields have been shown suppress the \hyd~abundance and 
thereby to prevent gas in ACHs from cooling below the virial temperature 
of the halo \citep[see][and references therein]{IVHRev20}. This threshold 
LW flux strength is commonly referred to as the critical flux or 
``$J_{\rm crit}$'' and has been typically found in hydrodynamic 
simulations to be in the range $10^{3-4}$ in the customary units 
$10^{-21}~{\rm erg~s^{-1}~cm^{-2}~Hz^{-1}~sr^{-1}}$. While this is 
orders of magnitude larger than the expected cosmological background 
\citep[e.g.][]{Dijkstra+08}, a collapsing halo near a particularly bright 
neighboring galaxy with recently-formed Pop III stars may be exposed to a 
such a flux \citep{VHB14b,Regan+17}; in this "synchronized collapse" 
scenario, if the two collapse within a short period of time -- of order 
a few Myr -- the second halo to cross the atomic cooling threshold may 
have \hyd--cooling entirely suppressed.

The presence of a super--critical flux has implications for the 
formation of massive seed black holes, $\sim 10^{(4-5)} M_\odot$;
rapid accretion in ACHs that remain near the virial temperature; 
these ``heavy seeds'' could help explain the existence of the earliest 
supermassive black holes, observed to have masses $\gsim 10^9 \msun$ 
at redshifts $z \gsim 6$ and as high as $\gsim 7.5$  
\citep{Fan+01,Fan+03,Morganson+12,Mazz+17,Wang+19,Yang+19,Wang+21}
These heavy seeds are commonly referred to as ``direct collapse'' black 
holes, though they're thought to form via an intermediary supermassive 
star phase \citep[e.g.][]{Haemmerle+18}.

Extensive work has been done to constrain the value of \jcrit~using 
hydrodynamic simulations of ACHs, which relies on detailed modeling 
of the \hyd~chemistry. Since the fraction of haloes exposed to a 
super--critical UV flux depends sensitively on \jcrit, even small 
changes in the photodissociation rate significantly alters the 
predicted prevalence of direct collapse halo candidates
\citep{Dijkstra+08,Ahn+09,Agarwal+12,DFM14,Chon+16}. 

\subsection{The ${\bf H_2}$ photodissociation rate}
Self--shielding by \hyd~occurs as the LW bands become optically thick
at column densities $\gsim 10^{13}~{\rm cm}^{-2}$, suppressing the 
photodissociation rate \citep[e.g.][]{DB96}. The optically--thick rate 
in general depends on the column density, gas temperature, rovibrational 
populations of \hyd \citep{WGH19}, and details of the incident spectrum
\citep{AK14,Sugimura+14,WGHB17}, and is prohibitively computational expensive 
to calculate on--the--fly in simulations, due to the large number of LW 
transitions. Simulations most often therefore implement a fitting formula 
to model the optically--thick rate and rely on local estimates of the 
column density (\citealt{WGHB11}, but see \citealt{Hartwig+15}).

In addition to self--shielding, absorption of LW photons by neutral hydrogen 
Lyman series resonances can decrease the rate of \hyd--photodissociation.
Processing of the cosmological UV background by HI in the pre--reionization 
IGM has been studied in detail \citep{HRL97,HAR00}; however, the effects of 
HI absorption {\it within} protogalactic halos has not previous been included 
in 3D simulations. Using one--zone models, \citet{WGH11} found that this 
HI--shielding of \hyd~can significantly decrease the LW photodissociation rate 
when the column density exceeds $N_{\rm HI}\gsim 10^{22}~\rm{cm^{-3}}$ and 
provided an analytic fit for the suppression factor $f_{\rm shield,HI}$.

In a study of the escape fraction of LW {\it out of} ACHs, 
\citet{Schauer+15} used the WH11 fitting formula to quantify the
effect of HI absorption of LW photons emitted stars within the halo. 
They found the LW escape fraction was reduced by a factor of three, 
owing to the large neutral column density. In a later study of more 
massive halos, $10^{7-8} M_\odot$, \citet{Schauer+17} found a 
significantly smaller effect, with escape fractions reduced by up to 
$\sim 29$ per cent due to HI absorption, possibly due to significantly 
more ionization by stellar clusters in the haloes resulting in lower 
neutral column densities. Nevertheless, these results point to the 
possible importance of HI--shielding of \hyd~in primordial ACHs 
irradiated by an external LW field.

In this study, we use the three--dimensional hydrodynamic simulation 
code \enzo~to test the effect of absorption by HI on \jcrit~in three 
UV--irradiated protogalaxies collapsing at z$\sim 10$. We use a modified
version of the WH11 fitting formula for HI--shielding of \hyd~that we 
updated to better fit non--ground state \hyd~rovibrational populations,
which become important at the temperatures and densities of gravitationally 
collapsing ACHs. Our modified fitting formula, obtained using data for 
the \hyd~rovibrational populations from {\sc cloudy}, is accurate to 
within $\sim 30$ per cent at $T=500-8000$K, $n=10^{0-5}~{\rm cm^{-3}}$, 
and column densities $N_{\rm HI} \lsim 10^{24}~cm^{-2}$,
and can be easily implemented in chemical models for future studies.

The rest of this paper is organized as follows. In \S~\ref{sec:Model} 
we provide details of our simulations and calculations of the HI--shielding; 
we discuss our results in \S~\ref{sec:Results} and conclude with a summary 
in \S~\ref{sec:Conclusions}.

\section{Numerical Modeling}
\label{sec:Model} 

\begin{figure*}
  \includegraphics[width=0.92\textwidth,trim={1cm 1.5cm 1cm 0cm,clip}]{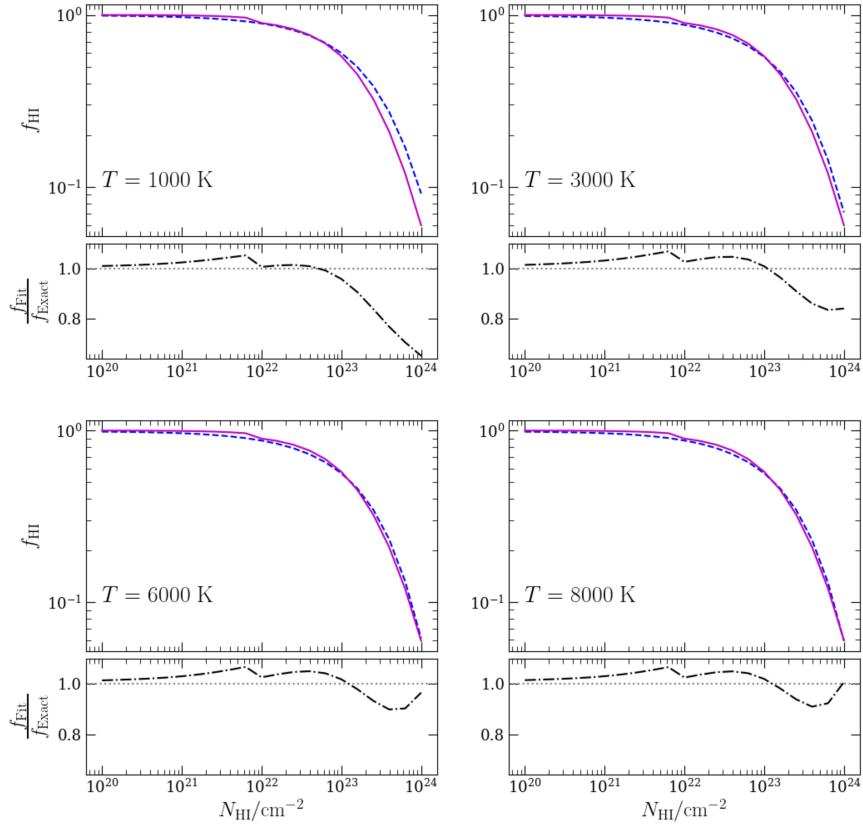}
\caption{The effect of shielding by HI on the H2 ~photodissociation rate is parameterized by a shielding factor 
      $f_{\rm HI}=k_{\rm diss}(n,T,N_{\rm H_2},N_{\rm HI})/k_{\rm diss}(n,T,N_{\rm H_2})$.
      The blue dashed lines show the HI--shielding factor from the full 
      calculation, $f_{\rm exact}$, with {\sc cloudy}--derived rovibrational 
      populations for ${\rm H_2}$. The magenta solid lines show our fit. 
      The black dot-dashed line in each lower panel shows the ratio of our fit 
      to the exact calculation. In order to isolate the HI fitting formula 
      accuracy from that for self--shielding, we calculate 
      $f_{\rm{fit}}$ as = $f_{\rm{HI,fit}} \times f_{\rm{H_2,exact}}$.
      All are shown for $\log (n / {\rm{cm^{-3}}})=3$, and 
      $\log(N_{\rm{H_2}}/{\rm{cm^{-2}}})=16$.}
    \label{fig:fshield}
\end{figure*}

\begin{figure}
  \includegraphics[width=0.4\textwidth,trim={1cm 1cm 3cm 1cm,clip}]{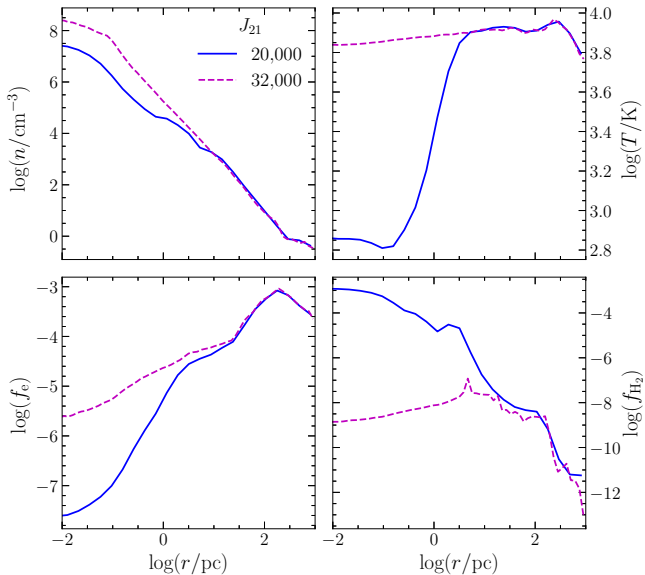}
  \caption{Spherically--averaged radial profiles for Halo B at the collapse
  redshift showing density (upper left), temperature (upper right), electron fraction 
  (lower left), and \hyd~fraction (lower right). Lyman-Werner fluxes are 
  $J_{21}=20,000$ (sub--critical; blue solid lines) and $J_{21}=32,000$ 
  (super--critical; magenta dashed lines). The radial distance is measured in physical units.}
  \label{fig:profiles}
\end{figure}
\begin{figure*}
  \includegraphics[width=0.9\textwidth,trim={0cm 2cm 0cm .5cm,clip}]{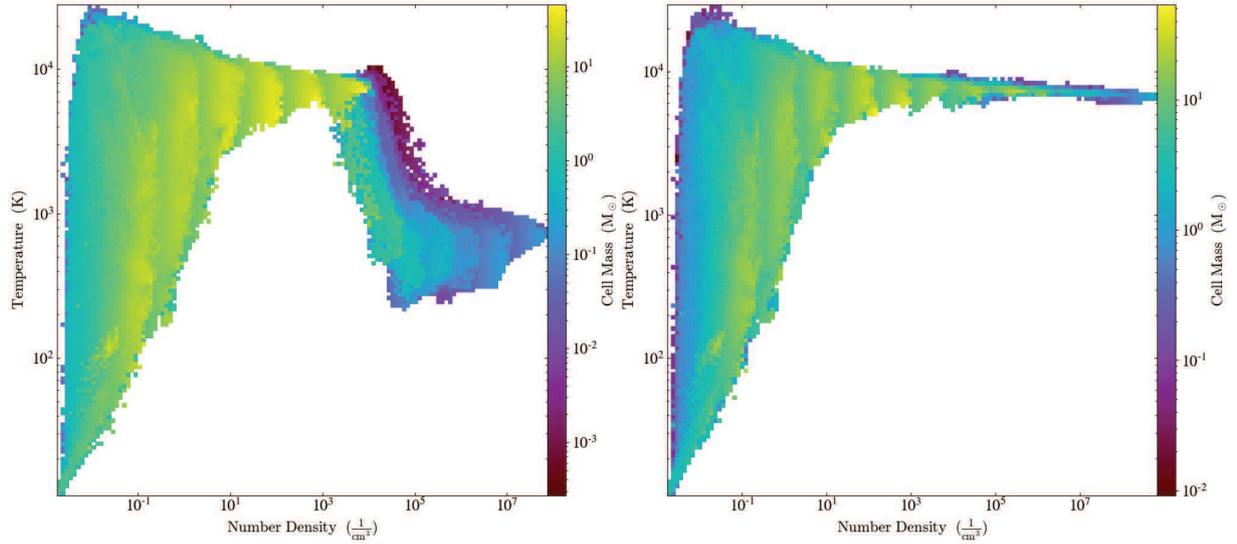}
  \caption{Phase plots showing Halo B at the collapse redshift with 
sub--critical flux (left) and super--critical flux (right).}
  \label{fig:phaseplots}
\end{figure*}

\begin{figure}
\centering
 \includegraphics[width=0.8\columnwidth,trim={3cm 8cm 3cm 8.5cm},clip]{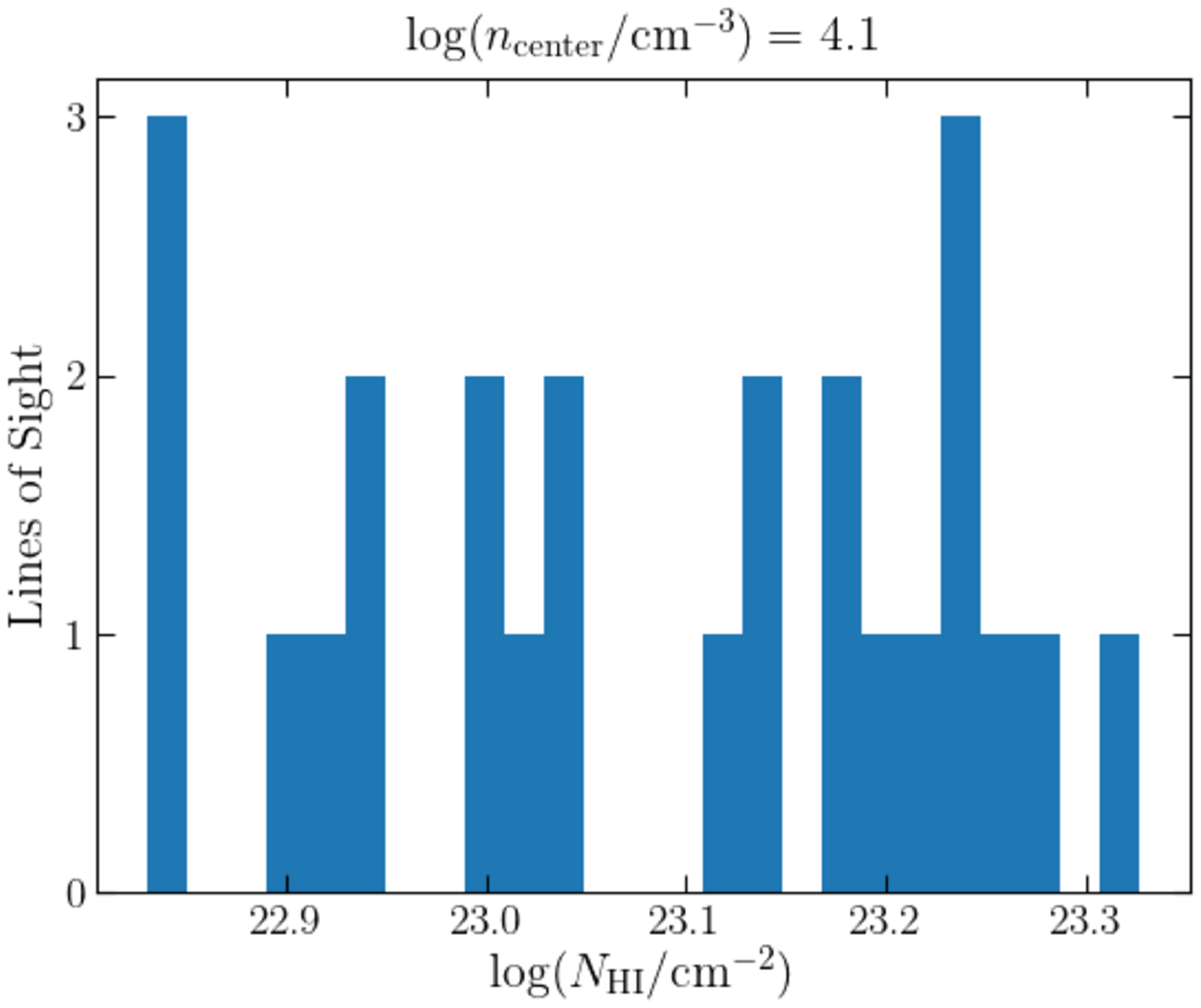}
  \caption{Histogram of HI column densities along lines of sight from the center of 
  Halo B just before cooling occurs (sub--critical flux). Column densities exceed the 
  $N_{\rm{HI}} = 10^{22}\ \rm{cm}^{-2}$ threshold, above which HI--shielding of 
  \hyd~becomes significant.}
  \label{fig:histogram}
\end{figure}

\subsection{Simulations}
\label{subsec:simulations}
We run simulations of three atomic cooling halos using \enzo, a publicly-available three-dimensional adaptive mesh refinement (AMR) hydrodynamic code \citep{Bryan+14}. 
Initial conditions for a box $1 h^{-1}$ Mpc on a side 
and $128^3$ root grid were generated using {\sc music} \citep{MUSICMethod11}. 
In order to select haloes for higher resolution “zoom--in” simulations, we 
performed an initial dark-matter only \enzo\ run from $z=99$ to $z=10$. 
We used the {\sc rockstar} halo finder package \citep{RockstarMethod13} to 
identify a halo with mass above the atomic cooling threshold at $z=10$; we then 
added three additional levels of refinement using nested grids which enclose 
the Lagrangian volume for the halo of interest, yielding an effective $1024^3$ 
resolution and dark matter particle mass $\sim 85 \msun$. 

Each halo is then run with $+$ DM ``zoom-in'' simulations initialized at 
$z=99$ to a maximum refinement level of 18, which results in the highest 
resolution regions having a minimum cell size of $.0298h^{-1}$pc. Additional 
refinement is added each time the baryon or dark matter mass exceeds four times 
that of the most refined cell. We also imposed that the local Jeans length is 
resolved by at least 16 cells to prevent spurious fragmentation \citep{Truelove+97}. 

We utilize the 9--species non--equilibrium primordial chemistry network within 
\enzo~to model the chemical evolution of the gas. The cooling function from 
\citet{Galli+Palla+98} is implemented to model the radiative cooling by \hyd. 
Several of the reaction rate calculations have been modified in the \enzo~chemistry code as described in \citealt{WGHB21} (see their Appendix A for details). 
For \hyd~self--shielding, we use the local column density from the ``Sobolev--like''
method described in WGHB11 and their fitting formula for the optically--thick 
rate.\footnote{This fit has since been updated by \citet{WGH19} to account for 
non-ground state rovibrational populations; however, using the updated fit would 
not affect our conclusions, since we are interested here in the change in 
\jcrit~due to HI--shielding, rather than the precise value of the critical flux.} We assume a blackbody incident radiation field at temperature $T=10^{5}$ K.

In order to determine the impact of HI--shielding, we run realizations of each 
halo to determine the value of \jcrit~first with \hyd~self-shielding only, using
the Newton-Raphson method to find \jcrit, and then subsequently run each with 
HI--shielding included, using our new fitting formula \S~\ref{subsec:fitting}. 

We use the publicly-available package {\sc YT} \citep{YT10} for simulation 
data analysis and visualization\footnote{yt-project.org}. Throughout, 
we adopt the cosmological parameters from the Planck 2018 collaboration 
\citep{Planck18}, $\Omega_{\rm m} = 0.315$, $\Omega_\Lambda=0.685$, 
$\Omega_b= 0.0493$, $h= 0.674$, $\sigma_8 = 0.811$, and $n = 0.965$. 

\subsection{HI--shielding of ${\bf H_2}$}
\label{subsec:fitting}

In order to find the exact optically--thick photodissociation rate, 
we use the method described in detail in WGH19 and summarized briefly 
here. The rate calculation includes contributions from LW transitions originating in the 301 bound rovibrational levels of the electronic 
ground state. We use the spectral synthesis code {\sc cloudy} \citep{Ferland+17} to model the \hyd~rovibrational populations at $T=(500-8000)$K, densities $T=10^{(0-5)}~{\rm cm^{-3}}$, $N_{\rm HI} = 10^{(20-25)}~{\rm cm^-2}$, and $N_{\rm H2} = 10^{(14-17)}~{\rm cm^-2}$. 
The fractional populations are then input in the rate calculation for 
each density and temperature combination. 

In order to determine the impact of HI--shielding, the rate is calculated 
with and without HI Lyman series absorption for each ${\rm n,T,N_{H2}}$. 
We define the dimensionless HI shield factor as: 
\begin{equation}
f_{\rm sh,HI} = \frac{k_{\rm diss}(n{\rm,T,N_{HI},N_{H2}})}
{k_{\rm diss}(n,{\rm T,N_{H2}})}.
\end{equation}
In order to develop our fit, we began with the form used in WH11, 

\begin{equation}
  f_{\rm sh,HI} = \frac{\chi}{(1+x)^{\delta}}\exp{(-\alpha x)}
\label{eq:fshieldHI}
\end{equation}

which was used in that study for photodissociation of \hyd~in the 
ground rovibrational state only; we modified the parameters using the 
downhill simplex method {\sc amoeba} provided in Numerical Recipes.

Here, $x = N_{\rm HI} / \zeta$ and our best fit parameters are: 
\begin{gather*}
\noindent \alpha = 1.45 \times 10^{-1} \\
 \delta = 1.5 \\
\noindent \zeta = 2.85 \times 10^{23} \rm{cm^{-2}} \\
\chi = \left\{
    \begin{tabular}{ll}
      $1$, & ${\rm N_{HI} < 10^{22}~cm^{-2}}$ \\
      $0.95$, & ${\rm N_{HI} \ge 10^{22}~cm^{-2}}$ \\
    \end{tabular}
  \right. 
\end{gather*}

Figure \ref{fig:fshield} shows the new HI-shielding factor fit, $f_{\rm fit}$
(Equation \ref{eq:fshieldHI}), and the exact shielding factor from the full 
calculation, $f_{\rm exact}$, at fixed $\log (n / {\rm{cm^{-3}}})=3$, 
$\log(N_{\rm{H_2}}/{\rm{cm^{-2}}})=16$ and a range of temperatures.
In the lower part of each panel is the ratio between $f_{\rm fit}$ 
and $f_{\rm exact}$. The fitting formula for the shielding function of 
\hyd~by HI is robust in the range of temperatures studied here, $500-8000$K 
and is accurate to within a factor of two for column densities $10^{20-24}~{\rm cm^{-2}}$.

\section{Results}
\label{sec:Results}
\begin{table}
  \begin{center}
  \caption{Critical fluxes with and without HI--shielding in $J_{21}$ units. 
    Virial masses and collapse redshifts indicated for $J<$\jcrit~runs with HI--shielding.}
  \label{tbl:Jcrit}
  \begin{tabular*}{0.47\textwidth}{@{\extracolsep{\fill}}c c c c l l}
  \hline\hline
          Halo & $M/10^7 M_{\odot}$ & $z_{\rm{coll}}$ & $T_{vir}/\rm{K}$ 
          & \jcrit$/10^3$ & \jcrit$_{\rm{,HI}}/10^3\
$ \\
    \hline
  ~A & $2.8$ & $13.0$ & $8,296$ & ~$11 $ & ~$22$\\
    ~B & $8.2$ & $10.9$ & $14,418$ & ~$17 $ & ~$32$\\
  ~C & $2.4$ & $18.0$ & $10,231$ & ~$10 $ & ~$16$\\
    \hline\hline\\
  \end{tabular*}
  \end{center}
  \vspace{-6mm}
  \end{table}

Figure \ref{fig:profiles} shows spherically--averaged radial profile of density, 
temperature, electron fraction, and \hyd~fraction for Halo A at the collapse 
redshift. Results for both sub--critical ($J_{21}=20,000$) and super--critical 
($J_{21}=32,000$) LW fluxes are shown. Our halos follow the well--known behavior 
of ACHs cooling in the presence of a photodissociating flux: with $J<J_{\rm crit}$, 
the \hyd~fraction in the halo's dense core reaches $\sim 10^{-3}$, the standard 
``freeze--out'' value \citep{OH02}; \hyd~cooling is efficient and the gas temperature 
falls below $10^3$K in the dense core. Irradiation by a super--critical flux results 
in a suppressed \hyd~fraction, $\sim 10^{-7}$, and the temperature remains at  
$T_{\rm vir} \sim 10^4 \rm{K}$ throughout the halo.

To determine \jcrit, we run the zoom-in simulations with varied levels of incident 
$J_{\rm LW}$ to find minimum flux that suppress \hyd--cooling and prevents cooling 
below the virial temperature. The resulting \jcrit~values for each of the three 
haloes are listed in Table \ref{tbl:Jcrit}. {\it We find that the critical flux 
with HI--shielding of \hyd\ is $\sim$60-100 percent larger than without HI--shielding.}
The $J_{\rm crit,21}$ values without HI--shielding for these halos are within the 
range $(10-17) \times 10^3$, comparable to those found in previous studies, and 
with HI--shielding $J_{\rm crit,21}$  are within the range $(16-32) \times 10^3$.
Figure \ref{fig:phaseplots} shows phase plots for both sub--critical
and super--critical fluxes for Halo B at the collapse redshift.

The increase in \jcrit\ with HI--shielding of \hyd indicates that the neutral column
densities are sufficient in these ACHs for Lyman series absorption to be important,
which occurs at ${\rm N_{HI} \gsim 10^{22}~cm^{-2}}$ (see Figure \ref{fig:fshield}).
In order to verify this, we find the column densities at the critical density\footnote{above 
which, collisional dissociation dominates the total \hyd~destruction rate} by summing along 
sightlines extending from the densest point of the halo out to a radius of $100$ pc. 
We show a histogram of 25 such sightlines in Figure \ref{fig:histogram} at the
time when the gas in the core has reached at the critical density, just before 
runaway cooling occurs. (Results are shown here for Halo B and those from the 
other two halos are similar.We see that indeed the neutral columns have reached 
the threshold at which HI--shielding becomes significant.

\section{Conclusions}
\label{sec:Conclusions}

In this study, we examined the impact of HI--shielding of \hyd~on the thermal 
evolution of protogalactic atomic cooling haloes exposed to photodissociating 
UV radiation using three--dimensional hydrodynamic simulations. We find that 
incorporation of HI--shielding raised the value of the critical flux to suppress 
\hyd--radiative cooling, \jcrit, by $\sim 60 - 100\%$ in the three ACHs we 
studied. This increase may have important implications for the predicted number 
of candidate halos that could seed massive black holes at $z\sim10$ via direct 
collapse, which is sensitive to the critical flux.

We used an updated fitting formula to model the suppression of \hyd~photodissociation 
by HI, which can be used in future simulations. The modified fitting function is 
accurate to within $\sim 30$ per cent at $T=500-8000$K, $n=10^{(0-5)}~{\rm cm^{-3}}$ 
and $N_{\rm HI} = 10^{(20-24)}{\rm cm^{-2}}$.

\section*{Acknowledgments}
We thank Zolt\'{a}n Haiman and S. Peng Oh for helpful discussions 
during the course of this work. Meredith Neyer acknowledges funding from an 
Edison STEM summer research program grant at University of California Santa 
Barbara. This material is based upon work supported by the National Science
Foundation under Award No. 1903935. This work used the Extreme Science and 
Engineering Discovery Environment (XSEDE; allocation TG-PHY200043), which is 
supported by National Science Foundation grant number ACI-1548562.

\section{Data availability}
The data underlying this paper will be shared on reasonable request
to the corresponding author.

\bibliography{paper}
\label{lastpage}

\end{document}